\documentclass[10pt,letterpaper]{article}
\usepackage{opex3}
\usepackage{epsfig}
\usepackage{amsmath}
\usepackage{amssymb}
\usepackage{mathrsfs}

\begin{document}

\title{Levitated droplet dye laser}

\author{H. Azzouz,$^{1}$ L. Alkhafadiji,$^2$ S. Balslev,$^1$\\ J. Johansson,$^3$ N.~A. Mortensen,$^1$
S. Nilsson,$^2$ and A. Kristensen$^1$}

\address{$^1$MIC -- Department of Micro and Nanotechnology, Nano$\bullet$DTU,\\ Technical University of Denmark, Building 345east, DK-2800 Kongens Lyngby, Denmark \\
$^2$Pure and Applied Biochemistry, Center for Chemistry and Chemical Engineering,\\ Lund University, P.O. Box 124, 221 00 Lund, Sweden \\
$^3$AstraZeneca, R\&D M{\"o}lndal, Analytical Development,
M{\"o}lndal, Sweden}

\email{ak@mic.dtu.dk}

\date{\today}

\begin{abstract}
We present the first observation, to our knowledge, of lasing from
a levitated, dye droplet. The levitated droplets are created by
computer controlled pico-liter dispensing into one of the nodes of
a standing ultrasonic wave (100 kHz), where the droplet is
trapped. The free hanging droplet forms a high quality optical
resonator. Our 750 nL lasing droplets consist of Rhodamine 6G
dissolved in ethylene glycol, at a concentration of 0.02 M. The
droplets are optically pumped at 532 nm light from a pulsed,
frequency doubled Nd:YAG laser, and the dye laser emission is
analyzed by a fixed grating spectrometer. With this setup we have
achieved reproducible lasing spectra in the visible wavelength
range from 610 nm to 650 nm. The levitated droplet technique has
previously successfully been applied for a variety of
bio-analytical applications at single cell level. In combination
with the lasing droplets, the capability of this high precision
setup has potential applications within highly sensitive
intra-cavity absorbance detection.
\end{abstract}



\section{Introduction}

Optical micro cavities in general are receiving considerable
interest due to their size and geometry dependent resonant
frequency spectrum and the variety of
applications~\cite{Vahala:2003}. Micro-droplets constitute an
interesting medium for micro-cavity optics in terms of e.g. lasing
emission properties~\cite{Qian:1986b} and manifestation of chaotic
wave dynamics~\cite{Mekis:1995,Nockel:1997}. Combined with
ultrasonic levitation techniques~\cite{Lierke:1996} micro-droplets
may also provide an interesting environment for analytical
chemistry~\cite{Santesson:2000} including intra-cavity
surface-enhanced Raman
spectroscopy~\cite{Santesson:2003,Symes:2004}.

Lasing in freely falling droplets was first reported by Qian
\emph{et al.}~\cite{Qian:1986b} which stimulated a significant
interest in the optical properties of droplets, see e.g.
Datsyuk~\cite{Datsyuk:2001a} and references therein. Parallel to
this there has been a significant attention to ultrasonic
levitation~\cite{Lierke:1996} from the chemical
community~\cite{Santesson:2000} allowing for studies of e.g.
protein crystallization~\cite{Santesson:2003a}. Combining
ultrasonic levitation with the optical properties of droplets
holds great promise for highly sensitive intra-cavity absorbance
detection system with prospects for single-molecule
detection~\cite{Symes:2004}. However, such applications rely
heavily on both reproducible loading of droplets and subsequent
reproducible generation of laser emission spectra by external
pumping.

In this paper we present the first observation, to our knowledge,
of reproducible lasing from levitated 750~nl dye droplets. Droplets are
optically pumped at 532 nm by a pulsed, frequency doubled Nd:YAG
laser and emission is analyzed by fixed grating spectrometry. The
levitated droplet constitutes a highly sensitive intra-cavity
absorbance detection system with prospects for single-molecule
detection~\cite{Symes:2004} and the possibility for
computer-generated on-demand droplets holds great promise for
applications in high-throughput chemical analysis.

\begin{figure}[b!]
\begin{center}
\epsfig{file=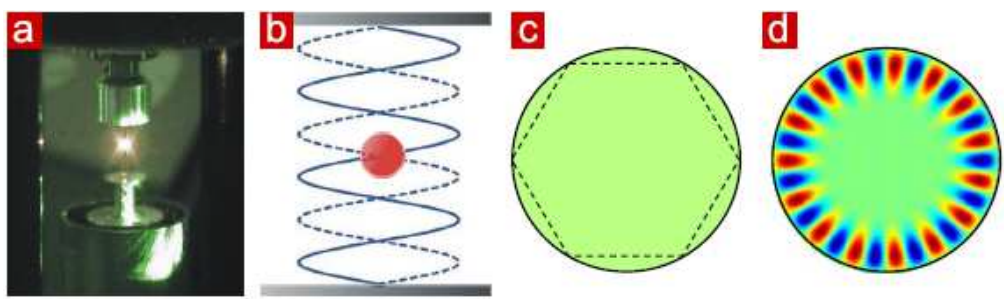, width=1\columnwidth, clip}
\end{center}
\caption{a) Photograph of a lasing levitated micro-droplet. b)
Schematics of ultrasonic field with the micro-droplet being
trapped at a node in the ultrasonic field. c) Schematics of
whispering-gallery modes in a (2D) spherical cavity. d) Numerical
example of a whispering-gallery mode in a (2D) spherical cavity.}
\label{fig1}
\end{figure}

\begin{figure}[t!]
\begin{center}
\epsfig{file=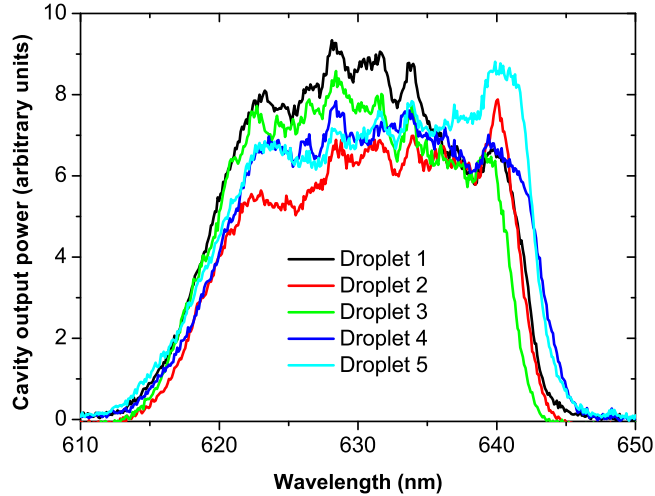, width=0.75\columnwidth, clip}
\end{center}
\caption{Reproducible lasing spectra from dye doped
micro-droplets. Each spectrum is obtained in a fixed setup by a
well-controlled loading of an EG droplet with a Rh6G dye which is
subsequently pumped above threshold, see Fig.~\ref{fig3}(b). The spectra are averaged over three pump pulses.} \label{fig2}
\end{figure}

\section{Experimental setup and results}

Ultrasonic levitation is a technique that facilitates the
performance of a variety of investigations on small volumes of
samples, i.e. liquid droplets and particles. It suspends the
object levitated in the nodal point of an ultrasonic standing
wave, see Fig.~\ref{fig1}(b). The technique was introduced in the
1930's and does not rely on any specific properties of the sample
except size and mass. The method has been used extensively in
bio-analytical and analytical chemistry applications, see
e.g.~\cite{Santesson:2000,Santesson:2004} and references therein.

The ultrasonic levitator~\cite{levitator} consists of an
ultrasonic transducer and a solid reflector supporting standing
waves, see Fig.~\ref{fig1}(a) and (b). In this work the levitator is
operated at a frequency of $\Omega_{\rm vib}/2\pi\sim 100\,{\rm
kHz}$ corresponding to a wavelength of $\Lambda\sim 6$~mm. The
levitator can hold particles of radius $a< \Lambda$ and droplets
with $a\sim \Lambda/6$ require a minimum of ultrasonic power.
Large droplets may be deformed by the body forces (gravity and
ultrasonic pressure gradients), which in practice limits the
droplet size to $a<\xi$ where the capillary length $\xi$ is in the
millimeter range for a water droplet. Furthermore, the droplet
shape may also be spherically deformed by applying a large
ultrasonic pressure amplitude.

In our experiments we used Rhodamine 6G (Rh6G) laser dye dissolved
in ethylene-glycole~(EG). The liquid sample was placed in a nodal
point of the levitator by means of a computer controlled
piezo-electric micro dispenser~\cite{Fiehn:1997}. A droplet with a
total volume of $V=(4\pi/3)a^3\sim 750\,{\rm nl}$ was formed by
repeated addition of pL drops.

The levitated dye droplet was optically pumped by a pulsed,
frequency-doubled Nd:YAG laser ($\lambda=532\,{\rm nm}$) with a
pulse length of 5~ns and a repetition rate of 10~Hz. The light
emitted from the micro-droplet was collected by an optical fiber
placed at an angle of approximately 50 degrees relative to the
pump laser beam. The emission was analyzed in a fixed-grating
spectrometer with a resolution of 0.15~nm.

\begin{figure}[b!]
\begin{center}
\epsfig{file=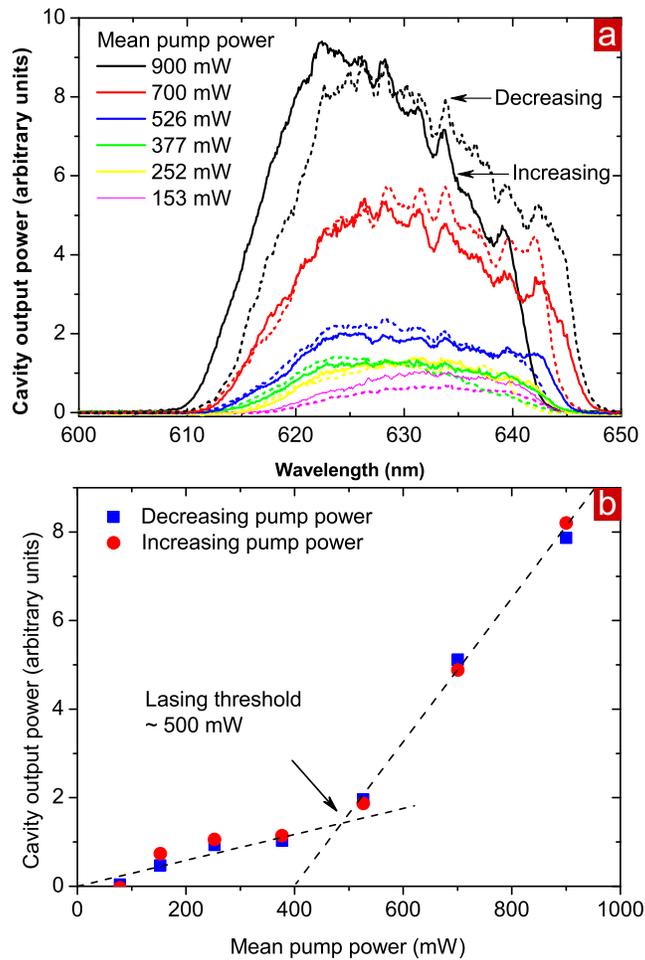, width=0.75\columnwidth, clip}
\end{center}
\caption{a) Cavity output power for increasing and decreasing
average pump power. Each spectrum is obtained in a fixed setup by
pumping an EG droplet with a Rh6G dye. The pump power is first
increased from zero up to level around $1000\,{\rm mW}$ (dashed
curves) and subsequently again lowered (solid curves). The spectra are averaged over three pump pulses. b) Cavity
output power versus mean pump power. The dashed lines are guides
to the eyes indicating a lasing threshold of around $500\,{\rm
mW}$ in the mean pump power.} \label{fig3}
\end{figure}

Evaporation and dye bleaching could hinder the applicability of
the dye droplet as a lasing device. In order to minimize these
effects, we used a measurement scheme, where nominally identical
droplets are loaded consecutively, and each droplet is only pumped
with 100 pulses (corresponding to a duration of 10~s) from the
Nd:YAG laser, before it is replaced by the next droplet. We have not systematically investigated the performance of the dye droplet lasers for more than 100 pulses.

In Fig.~\ref{fig2} we show 5 emission spectra obtained from
normally identical $V=750$~nl EG droplets, with a Rh6G
concentration of $2\times 10^{-2}\,{\rm mol/L}$. The laser was
pumped above the threshold. The observed variations in output
power are attributed to fluctuations in the pump power.

In a second measurement series we demonstrate the lasing action of
the dye droplet where consecutively loaded droplets are pumped at
different average pump power. In Fig.~\ref{fig3}(a) we show the
measured spectra and in Fig.~\ref{fig3}(b) we show the dye droplet
output power versus pump power. In the measurement sequence the
pump power was first increased from 150~mW to 900~mW and
subsequently decreased again. The reproducibility of the obtained
spectra and the lasing threshold is seen from panels a and b,
respectively.


\section{Discussion}

In the following we briefly address the optical and mechanical
modes, to assess the influence on the optical performance of
levitated droplet dye lasers and their applications for
intra-cavity sensing.

\subsection{Optical modes}

For a spherical resonator the whispering-gallery modes (WGMs) are
characterized by their radial quantum number $p$ as well as by
their angular momentum quantum number $\ell$ and the azimuthal
quantum number $m$ which can have $(2\ell+1)$ values, i.e.
$\omega_{p\ell}$ has a $(2\ell+1)$ degeneracy~\cite{Symes:2004}.
For the lowest-order radial modes ($p=1$), see Fig.~\ref{fig1}(d),
the resonances are to a good approximation given by
$n\chi\sim\ell$ where $\chi=ka$ is the so-called dimensionless
size parameter, $n$ is the refractive index, $a$ is the radius of
the droplet, and $k=\omega/c=2\pi/\lambda$ is the free-space wave
vector. The modes are thus equally spaced in frequency and the
corresponding spacing in wavelength is~\cite{Symes:2004}
\begin{equation}
\Delta \lambda = \left|\frac{\partial\lambda}{\partial
\chi}\right|\Delta\chi \simeq \frac{\lambda^2}{2\pi
a}\frac{\tan^{-1}(n^2-1)^{1/2}}{(n^2-1)^{1/2}},
\end{equation}
where the last fraction indeed has a $1/n$ asymptotic dependence
for a large refractive index, $n\gg 1$. For the EG droplets in the
experiments the corresponding mode spacing is of the order
$\Delta\lambda\sim 0.1$~nm which is not resolved in our present
experiments. However, $V=100$~pL droplets ($a\sim 30\,{\rm \mu
m}$) have been achieved in different experiments by operating the
levitator at 100~kHz. This would increase the mode spacing to
$\Delta\lambda\sim 1$~nm.

\subsection{Droplet shape and mechanical modes}

The shape of the droplet may be understood from Gibbs free-energy
arguments where one naturally defines the characteristic capillary
length $\xi=\sqrt{\gamma/\rho g}$ with $\gamma$ being the surface
tension, $\rho$ is the liquid mass density, and $g$ is magnitude
of the gravitational field~\cite{Landau:1987}. A droplet with a
characteristic size $a\ll \xi$ will have a shape strongly
determined by the surface tension and a spherical shape will then
obviously minimize the Gibbs free energy. For a free water droplet
in air $\xi \sim 2.7\,{\rm mm}$. In our experiments, where $a\sim
500\,{\rm \mu m}$, the spherical shape is not perturbed
significantly by body forces.

We emphasize that in principle the levitated droplet is a
complicated dynamical system, see e.g.~\cite{Yarin:2002}. However,
as analyzed already by Rayleigh in 1879~\cite{Landau:1987} the
vibrational spectrum of a liquid droplet originates from two
classes of modes; surface-tension driven surface modes and
compression-driven compressional modes. Since the liquid can be
considered incompressible the latter class of modes is in the
high-frequency regime while the low-frequency response is due to
surface-shape oscillations conserving the volume. The surface
vibrational modes are similar to the optical WGMs and are
characterized by their angular momentum quantum number $\ell_{\rm
vib}$. For low amplitude oscillations Rayleigh found
that~\cite{Landau:1987}

\begin{equation}
\omega_{\rm vib}=\sqrt{\frac{\gamma \ell_{\rm vib}(\ell_{\rm
vib}-1)(\ell_{\rm vib}+2)}{\rho a^3}},\ell_{\rm vib}=2,3,4,\ldots
\end{equation}
A droplet of a given radius $a$ can thus not be vibrationally
excited by the ultrasonic pressure field at frequencies
$\Omega_{\rm vib}<\sqrt{8\gamma/\rho a^3}$. For a driving
frequency of 100 kHz this implies that water droplets of radius
below $10\,{\rm \mu m}$ are not vibrationally excited.

\subsection{Prospects for intra-cavity sensing}

The prospects for intra-cavity sensing in liquid dye droplets
correlate strongly with the optical cavity $Q$ factor. The WGMs
each have a resonant frequency $\omega$ with a width
$\delta\omega=1/\tau$ where $\tau$ is the lifetime of a photon in
the mode. The corresponding quality factor $Q=\omega/\delta\omega$
of WGMs is determined by several factors including intrinsic
radiative losses originating from the finite curvature of the
droplet surface $Q_{\rm rad}$, absorption due to the cavity medium
$Q_{\rm abs}$, and broadening of resonances due to vibrational
interaction with the ultrasonic field $Q_{\rm vib}$, i.e.

\begin{equation}
Q^{-1}=Q_{\rm rad}^{-1}+Q_{\rm abs}^{-1}+Q_{\rm vib}^{-1}.
\end{equation}
For the radiative loss $Q_{\rm rad}$ increases close to
exponentially with the size parameter $\chi$ and for a refractive
index $n\sim 1.45$ we have $Q_{\rm rad}\sim 10^{5}$ and $Q_{\rm
rad}\sim 10^{12}$ for $\chi$ equal to 50 and 100,
respectively~\cite{Buck:2003}. In the case of bulk absorption we
have~\cite{Gorodetsky:1996}
\begin{equation}
Q_{\rm abs}=\frac{2\pi n}{\alpha\lambda}=\frac{\chi}{\alpha a}
\end{equation}
where $\alpha$ is the absorption coefficient of the cavity medium.

The broadening of the WGMs by interaction with the vibrational
modes is complicated, but we may immediately disregard the
high-frequency compressional modes leaving only the low-frequency
surface-tension driven modes for concern. Even though the
realistically attainable droplet ($a> 40\,{\rm \mu m}$) will be
vibrationally excited by the ultrasonic pressure field, the
influence of the disturbance could be suppressed by short pump
pulse operation. Finally, static, deformations of the cavity may
give rise to partial chaotic ray dynamics with a universal,
frequency independent, broadening of the WGM resonances which will
decrease $Q_{\rm rad}$~\cite{Mekis:1995,Nockel:1997}. A similar
vibration-induced decrease of $Q_{\rm rad}$ is expected.

\section{Conclusion}

In conclusion we have demonstrated a reproducible laser action
from an ultrasonically levitated laser dye droplet, when nominally
identical 750~nL droplets are consecutively loaded and optically
pumped by a pulsed frequency-doubled Nd:YAG laser. The present
droplets show reproducible multi-mode lasing. This system is
considered a potential candidate for intra-cavity sensing, and the
limitations induced by ultrasonic field were discussed.

\section*{Acknowledgments}
This work is supported by the Danish Technical Research Council
(grant no. 26-02-0064) and the Danish Council for Strategic
Research through the Strategic Program for Young Researchers
(grant no. 2117-05-0037). Financial support from the Swedish
Research Council (VR), Crafoordska Stiftelsen, Kungliga
Fysiografiska S\"{a}llskapet i Lund, Centrala
F\"{o}rs\"{o}ksdjursn\"{a}mnden, and the R. W. Johnson Research
Institute is gratefully acknowledged.

\end{document}